\documentclass[aps,pre,twocolumn,groupedaddress,11pt,tightenlines]{revtex4-1}%
\usepackage{graphicx,amsmath,amssymb,flushend}

\begin{document}

\title{Bistable dynamics of control activation in human intermittent control}
 
\author{Arkady Zgonnikov}
\email[]{arkady@u-aizu.ac.jp}

\author{Ihor Lubashevsky}
\email[]{i-lubash@u-aizu.ac.jp}
 
\affiliation{University of Aizu, Tsuruga, Ikki-machi, Aizuwakamatsu, Fukushima 965-8580, Japan} 

\begin{abstract}
When facing a task of balancing a dynamic system near an unstable equilibrium, humans often adopt intermittent control strategy: instead of continuously controlling the system, they repeatedly switch the control on and off. Paradigmatic example of such a task is stick balancing. Despite the simplicity of the task itself, the complexity of human intermittent control dynamics in stick balancing still puzzles researchers in motor control. Here we attempt to model one of the key mechanisms of human intermittent control, control activation, using as an example the task of overdamped stick balancing. In so doing, we focus on the concept of noise-driven activation, a more general alternative to the conventional threshold-driven activation. We describe control activation as a random walk in an energy potential, which changes in response to the state of the controlled system. By way of numerical simulations, we show that the developed model captures the core properties of human control activation observed previously in the experiments on overdamped stick balancing. Our results demonstrate that the double-well potential model provides tractable mathematical description of human control activation at least in the considered task, and suggest that the adopted approach can potentially aid in understanding human intermittent control in more complex processes.
\end{abstract}

\maketitle

\section{Introduction}
Humans face the task of balancing dynamic systems near an unstable equilibrium repeatedly throughout their lives, for example, when maintaining upright stance~\citep{loram2002direct}, carrying a cup of coffee~\citep{mayer2012walking}, or driving a car on a highway~\citep{lubashevsky2003rational}. The task of inverted pendulum (stick) balancing is an increasingly popular paradigm of studying human control behavior over unstable dynamic systems (see recent reviews in~\citep{asai2013learning, balasubramaniam2013control, milton2013intermittent}). Physics and control engineering have revealed a wide variety of continuous control strategies which can successfully stabilize upright position of inverted pendulum, including simple linear feedback~\citep{kwakernaak1972linear}, or, counterintuitively, high-frequency vibration applied to the pivot point of the stick~\citep{kapitza1951dynamic}. However, due to a number of physiological constraints, continuous control strategies either cannot be realized or are ineffective when applied by humans~\citep{loram2011human}. For this reason, much research has been aimed at understanding the mechanisms of discontinuous, or \textit{intermittent} control, which is often employed by humans in controlling unstable systems.

Intermittent control is characterized by repeated switching between periods of passive (control is off) and active (control is on) behavior of the controller. Previously it has been shown that intermittent control strategies are efficient and robust under presence of significant time delays and sensorimotor noise~\cite{milton2008unstable,milton2009time,milton2011delayed,loram2011human}. Moreover, numerous signs of intermittency have been detected in experimental data on human postural sway~\citep{loram2005active,loram2005human,bottaro2005body}, stick balancing on fingertip~\citep{cabrera2002onoff}, and, more recently, virtual stick balancing on the computer screen~\citep{bormann2004visuomotor,zgonnikov2014react}. Numerous attempts to model the dynamics observed in these experiments typically attribute the discontinuity of human control to the so-called "sensory deadzone"~\citep[e.g.][]{milton2009balancing,gawthrop2011intermittent,milton2013intermittent}. According to this hypothesis, the control remains switched off as long as the deviation of the controlled system from the desired state remains below certain threshold value. Whenever the deviation exceeds this threshold, the control is immediately switched on.

Models which incorporate threshold-driven control activation can explain much dynamics observed, e.g., in balancing of inverted pendulum and quiet standing. Still, the question remains open: How accurately does the threshold model describe the process of control activation in humans? Recent investigations of virtual overdamped stick balancing have revealed that control triggering at large deviations occurs much more frequently then predicted by threshold-driven activation~\citep{zgonnikov2014react}. Besides, some of the most peculiar phenomena observed in human control behavior still remain unexplained by the models based on threshold-driven control activation. Importantly, such phenomena include high occurrence of large fluctuations in human-controlled systems, which result, e.g., in falls during stick balancing or quiet standing~\citep{cabrera2012stick}. All these considerations suggest that in controlling even the simplest unstable systems humans may employ a somewhat more complex control activation mechanisms than assumed by conventional threshold-based models. 

In a recent attempt to elucidate the properties of control activation in humans, Zgonnikov et al. introduced the concept of intrinsically stochastic, noise-driven control activation~\cite{zgonnikov2014react}. The key hypothesis of this approach is that mathematical description of control activation should encompass stochasticity of human cognitive processes during decision when to start actively controlling the unstable system. The plausibility of this hypothesis has been confirmed by the simple ad hoc model mimicking the effect of noise-driven activation. Yet, the adequate mathematical formalism, which would allow one to model noise-driven control activation in different tasks, is still missing.

The aim of the present work is to develop explicit mathematical description of noise-driven control activation. The model we develop here describes competition between two cognitive states of the human operator, ``wait'' and ``act'', which are mapped onto the two states of a bistable dynamical system. The stochastic dynamics of control activation is captured via the notion of random walk in a double-well potential, which has proven its efficiency in modeling complex cognitive processes~\citep[e.g.][]{tuller1994nonlinear,moreno2007noise,vanrooij2013modeling}. Using a simple example of overdamped stick balancing, we demonstrate that the model reproduces the behavior of human operators observed experimentally. Based on the results of numerical simulations, we argue that the double-well approach can serve as a natural, easily adaptable framework for modeling mechanisms of control activation in diverse human-controlled processes.

\section{Human balancing of overdamped stick}
\begin{figure}
\centering
\includegraphics[width=0.5\columnwidth]{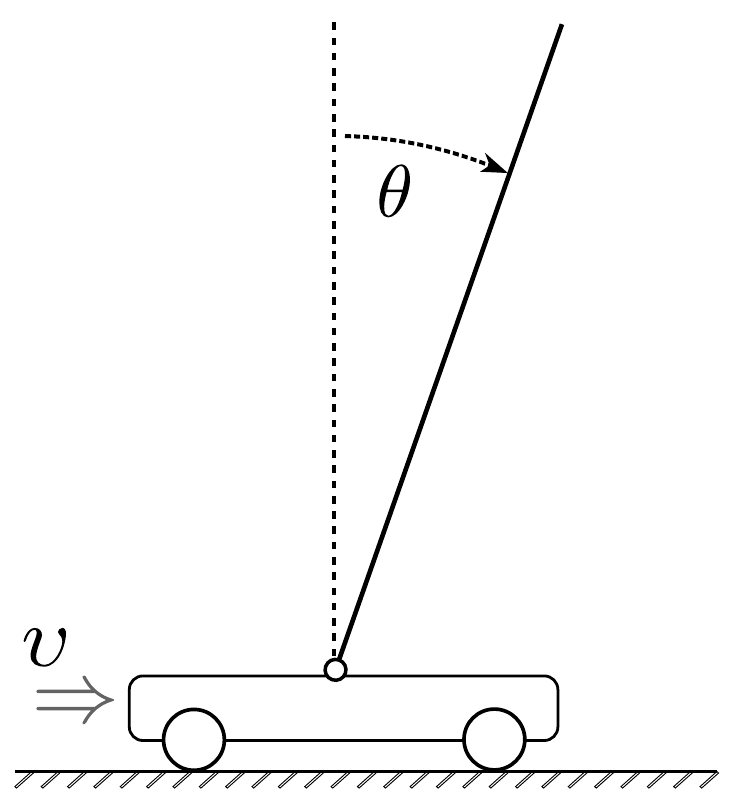}
\caption{
Inverted pendulum attached to a moving cart.
}
\label{fig:stick}
\end{figure}

Although we hypothesize the model developed here can be applied to virtually any human-controlled system, we exemplify our theoretical constructs using possibly the simplest unstable mechanical system, that is, overdamped stick balancing (Fig.~\ref{fig:stick}). The linearized dynamics of the mechanical system in terms of the stick angle~$\theta$ and cart velocity~$\upsilon$ (which is a control variable in this case) are described by the differential equation~(for derivation see~\cite{zgonnikov2014react})
\begin{equation}
\label{eq:stick}
 \tau_{\theta} \dot \theta = \sin \theta - \frac{\tau}{l} \upsilon \cos \theta,
\end{equation}
where $\tau_{\theta}>0$ is the intrinsic time scale of the stick motion, and $l>0$ is the length of the stick. 

\begin{figure}
\centering
\includegraphics[width=0.8\columnwidth]{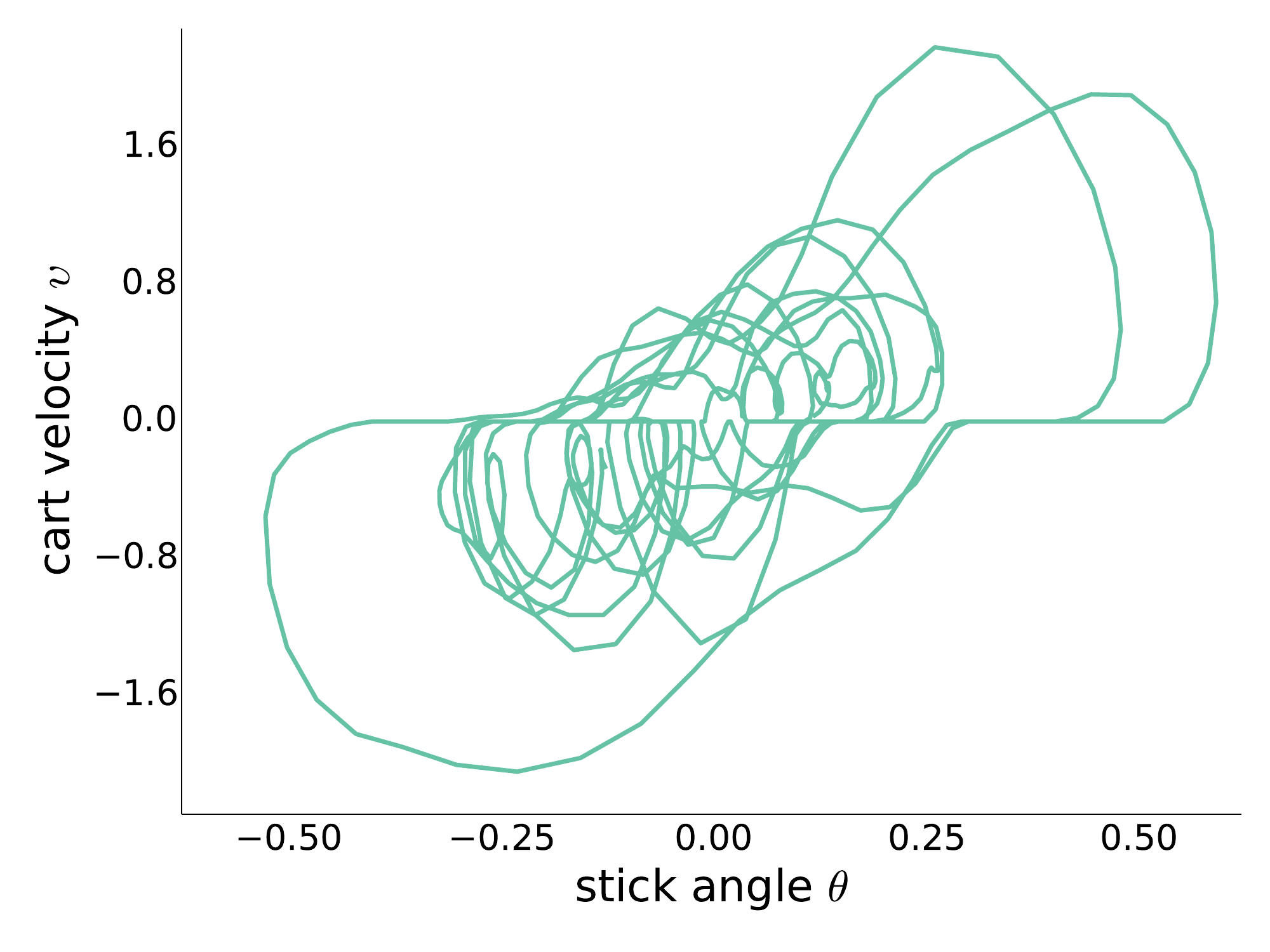}
\caption{Typical trajectory of overdamped inverted pendulum balancing by a human subject (based on the data reported in~\cite{zgonnikov2014react}). The trajectory reflects 15 seconds of balancing without stick falls.}
\label{fig:exp_trajectory}
\end{figure}

Experiments on human balancing of virtual overdamped stick on the computer screen revealed that the stick dynamics under human control are universal across the subjects and within a range of kinetic parameters of the stick~\citep{zgonnikov2014react}. The task for the subjects was to keep the stick upwards, whereas the cart could be moved arbitrarily within the limits of the computer screen. The experiments had shown that the subjects universally employed intermittent control strategy, with the cart velocity equal to zero for significant amount of time (Fig.~\ref{fig:exp_trajectory}). Importantly, the subjects often preferred to start correcting the stick position only when the deviation significantly exceeded the scale of the angle sensory deadzone. The distribution of \emph{action points} (stick deviations triggering reaction of the operator) was found to be considerably non-Gaussian, which suggests that factors other than sensory threshold determine the process of control activation. Without attempting to identify all such factors and model them in all their complexity, we will instead construct an integrated, phenomenological description of the control activation process.

\section{Double-well dynamics of control activation}
\subsection{Model description}
\begin{figure*}[!t]
\centering
\includegraphics[width=1.0\linewidth]{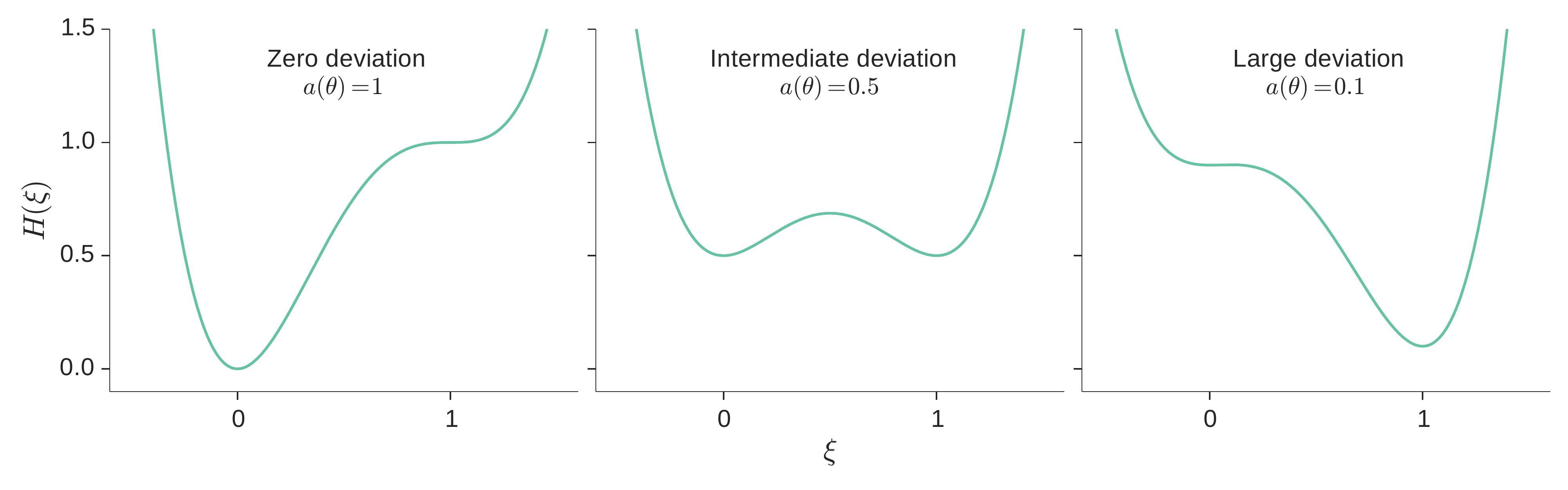}
\caption{
Double-well energy landscape $H(\xi)$ depending on deviation of the stick from the desired position (as characterized by $a(\theta)$).
}
\label{fig:potential}
\end{figure*}

In the case of overdamped inverted pendulum considered here, a model of the operator's behavior should specify how the control variable $\upsilon$ evolves in time given the state of the stick $\theta$. Within the paradigm of intermittent control, such a model should describe how the control is activated and how it is executed once activated. We have to emphasize, however, that the present model focuses mostly on the former issue. The latter problem, control execution, supposedly concerns the formalism of open-loop control, and, we believe, deserves detailed consideration elsewhere.

Appealing to the phase space extension approach~\citep{zgonnikov2014extended}, we consider the cart velocity $\upsilon$ an independent phase variable, so that its dynamics is defined by a separate differential equation, specifically
\begin{equation}
\label{eq:upsilon}
\dot \upsilon = \alpha l \theta \xi - \beta \upsilon, 
\end{equation}
where $\alpha$ and $\beta$ are non-negative constants, and $\xi$ is an order parameter describing the cognitive state of the operator in regards to the controlled system. The order parameter $\xi$ switches intermittently between the states $\xi=0$ and $\xi=1$, which are mapped onto the operators' two cognitive states, ``wait'' and ``act''. Consequently, in either the ``act'' ($\xi=1$) or ``wait'' ($\xi=0$) states the cart dynamics are effectively linear; it is the nonlinear mechanism of switching between these states that encompasses stochasticity and complexity.

In our model, the dynamics of $\xi$ is driven jointly by the deterministic and random forces. The deterministic dynamics are governed by the double-well potential energy landscape, where the configuration of the landscape is determined by the state of the controlled system (in our case, the stick angle $\theta$). The stochastic switching between the two wells is caused by a random force. Such dynamics can be described by the Langevin equation (in the It\^o interpretation)
\begin{equation}
\label{eq:xi}
\tau_{\xi} \dot \xi = -\frac{\partial H}{\partial \xi} + \sqrt{\varepsilon H} \zeta,
\end{equation}
where $\tau_{\xi}>0$ is the constant parameter defining the time scale of the switching process, $H(\xi,\theta)$ is the Hamiltonian shaping the energy landscape of the system (Fig.~\ref{fig:potential}), $\zeta$ is white noise, and $\varepsilon>0$ is the parameter regulating the noise intensity. 

We wish to underline that the model utilizes \emph{multiplicative} noise in Eq.~\eqref{eq:xi} to reflect the assumption that the level of uncertainty of the operator's cognitive state is low in unambiguous situations ($H\approx0$), and that this uncertainty increases with ambiguity of the current system state (which is quantified by $H$). Specifically, we assume that in the situations when the deviation $\theta$ is large, the operator's cognitive state is generally certain (that is, ``act''). The square-root dependence of the noise intensity on $H$ has been chosen for the sake of simplicity; in this case the form of the Langevin equation~\eqref{eq:xi} does not depend on the stochastic process interpretation. Namely, the difference between this equation written in the It\^o, Stratonovich, or H\"anggi-Klimontovich forms is reduced to a minor renormalization cofactor in the regular drift term (see e.g.~\cite{mahnke2009physics}).

The particular form of the Hamiltonian $H$ is chosen in a way that 
\begin{equation}
\label{eq:dHdxi}
\frac{\partial H}{\partial \xi}\propto\xi(\xi-1)(\xi-a(\theta)),
\end{equation}
where
\begin{equation}
\label{eq:a}
 a(\theta) = 1/(1+(\theta/\eta)^2).
\end{equation}
Parameter $\eta$ of the ansatz~\eqref{eq:a} characterizes the operator's perception, so that $a \approx 1$ if $|\theta| \ll \eta$ and $a \approx 0$ when $|\theta| \gg \eta$. Thus, when the stick deviation is large ($a \approx 0$), the energy landscape configuration is such that the ``act'' cognitive state ($\xi \approx 1$) can be expected. On the contrary, the ``wait'' state ($\xi \approx 0$) is most likely to be observed whenever the deviation is small ($a \approx 1$). Intermediate values of $\theta$ lead to bistable dynamics, so both the states become possible (Fig.~\ref{fig:potential}). 

We impose the following constraints on $H$
\begin{equation}
 \begin{split}
\label{eq:boundaries}
H|_{\xi=0,\,\theta=0}=0,\quad H|_{\xi=1,\,|\theta|\gg \eta}=0,
\\
H|_{\xi=0,\,|\theta|\gg\eta}=1,\quad H|_{\xi=1,\,\theta=0}=1.
\end{split}
\end{equation}
so that the random fluctuations diminish when the energy minimum $H=0$ is achieved at the states $\xi=0$ and $\xi=1$ in case the state of the controlled system is unambiguous ($\theta = 0$ and $|\theta|\gg\eta$, respectively). Conditions~(\ref{eq:dHdxi}--\ref{eq:boundaries}) yield
\begin{equation}
\label{eq:H}
H(\xi,a(\theta))= 3\xi^4-4(1+a)\xi^3 + 6a\xi^2+1-a.
\end{equation}

To simplify the model  analysis, we linearize Eq.~\eqref{eq:stick} around the upright position $\theta=0$ and rescale the time, stick angle, and cart velocity
\begin{equation*}
\label{eq:rescaling}
 t \to t \tau_{\theta},\quad \theta \to \theta \eta \quad \upsilon \to \upsilon \eta l /\tau_{\theta},
\end{equation*}
so that Eqs.~(\ref{eq:stick}--\ref{eq:xi}) are transformed into
\begin{equation}
\label{eq:model_rescaled}
 \begin{aligned}
  \dot \theta & = \theta - \upsilon, \\
  \dot \upsilon & = \gamma \theta \xi - \sigma \upsilon, \\
  \tau \dot \xi & = -\frac{\partial H}{\partial \xi} + \sqrt{\epsilon H} \zeta,  
 \end{aligned}
\end{equation}
where $\tau = \tau_{\xi}/\tau_{\theta}$, $\gamma = \alpha \tau_{\theta}^2$, $\sigma = \beta \tau_{\theta}$, and $\epsilon = \varepsilon/\tau_{\theta}$. In the rescaled variables the expression~\eqref{eq:a} takes form
\begin{equation}
\label{eq:a_rescaled}
 a(\theta) = 1/(1+\theta^2).
\end{equation}
while the form of the energy function~\eqref{eq:H} is conserved.

\begin{figure*}
\centering
\includegraphics[width=1.0\linewidth]{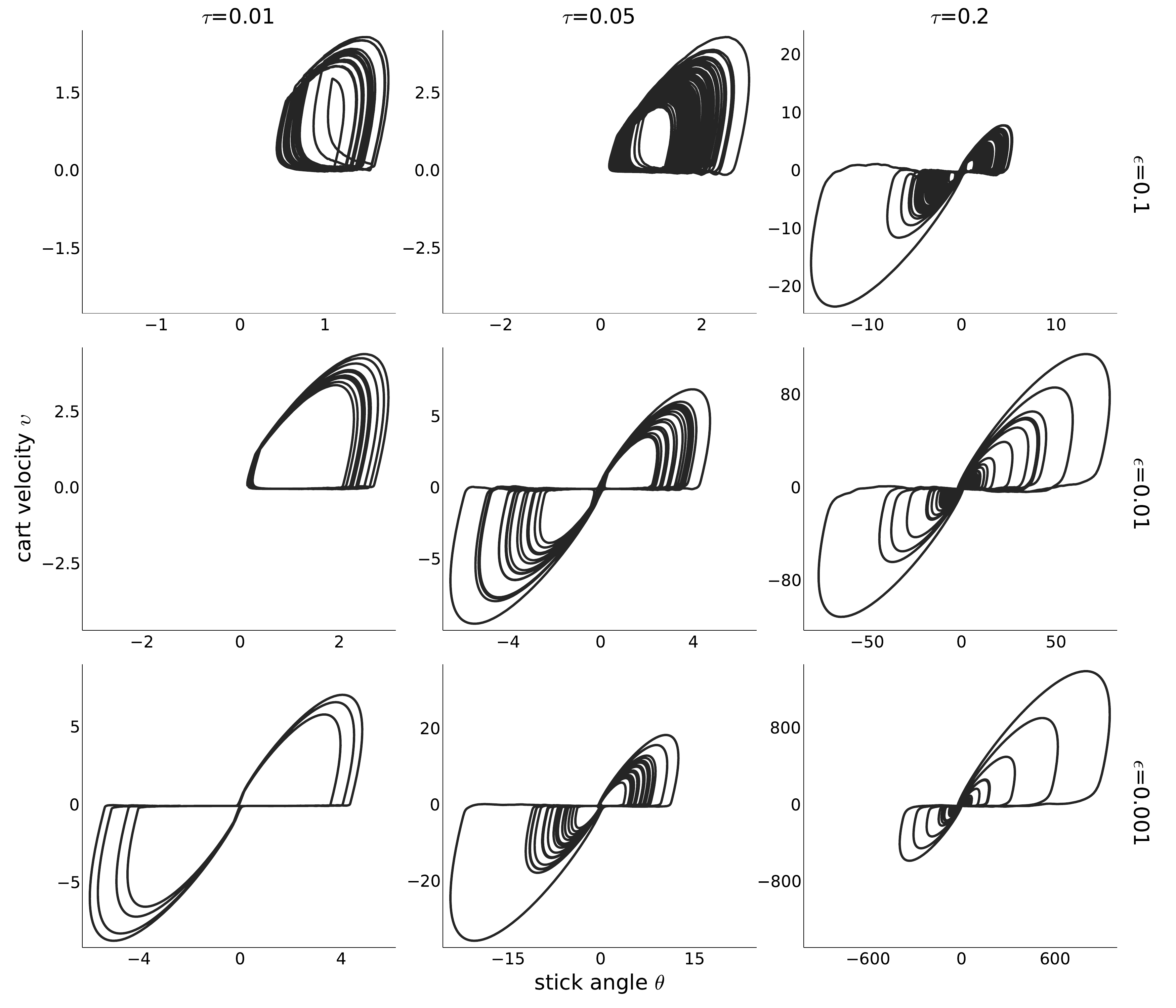}
\caption{
Projections of the model-generated trajectories on the $\theta-\upsilon$ plane. Each trajectory represents the system motion of duration $5000\tau$ time units.
} 
\label{fig:trajectories}
\end{figure*}

\subsection{Model analysis}
The model~(\ref{eq:H}--\ref{eq:a_rescaled}) has four parameters, $\tau$, $\epsilon$, $\gamma$, and $\sigma$. The feedback parameters $\gamma$ and $\sigma$ affect the system dynamics only when control is active ($\xi\approx1$). Under some weak assumptions (essentially, $\sigma >1$, $\gamma > \sigma$) the particular values of these parameters have no substantial effect on the system dynamics, so in what follows we assume $\sigma=3.5,\,\gamma=\sigma^2/2$, the values which were previously shown to be physically plausible for overdamped stick balancing~\citep{zgonnikov2014react}.

The other two parameters, $\tau$ and $\epsilon$, deserve close examination, because they define the very dynamics of the control activation process. The parameter $\tau$ denotes the time scale of the local dynamics of the operator's cognitive state $\xi$. More specifically, it defines how fast $\xi$ responds to changes of the energy landscape $H$, i.e., variations of the stick angle $\theta$. Smaller values of $\tau$ lead to faster transient process, and the other way around. The parameter $\epsilon$ affecting the noise intensity quantifies the overall level of uncertainty in the system. Specifically, for a given (fixed) configuration of the double-well potential~$\tilde{H}$, the rate of intermittent switching between the two wells increases with $\epsilon$.

To illustrate the dynamics of the model depending on $\tau$ and $\epsilon$, we have run numerical simulations using the stochastic Runge-Kutta scheme~\citep{rossler2005explicit} for the Ito-type stochastic process~(\ref{eq:H}--\ref{eq:a_rescaled}) with the discretization step $\Delta t = \tau/10$.

We consider the values of $\tau$ such that $\tau \ll 1$: to enable the controller to keep the stick upwards, the switching should apparently occur on time scales shorter than the time scale of the stick motion. The noise intensity parameter is also assumed to be small ($\epsilon \ll 1$); otherwise, the noise term loses its physical meaning. To be specific, we illustrate the system dynamics for $\tau \in \{ 0.01, 0.05, 0.2 \}$ and $\epsilon \in \{ 10^{-3}, 10^{-2}, 10^{-1} \}$.

The basic pattern of the system dynamics remains the same for all tested values of the system parameters~(Fig.~\ref{fig:trajectories}). The system initially perturbed by a small deviation of $\theta$ moves along the axis $\upsilon=0$, which represents the passive control phase ($\xi=0$). As the angle~$\theta$ increases, the energy landscape changes so that the transition to the active phase~($\xi=1$) becomes increasingly probable. This transition is induced by noise, so it occurs at probabilistically determined angle. Once the transition $\xi:\,0\to1$ occurs, the corrective feedback comes into action, and the stick is returned to the vicinity of the upright position. However, when the time scale $\tau$ is short and the noise intensity $\epsilon$ is high (e.g., top left frame in Fig.~\ref{fig:trajectories}), the control is always deactivated $\xi:\,1\to0$ before the stick reaches the vertical position (i.e., only undershooting is observed). In other tested cases, both the under- and overshooting corrections are implemented, which more resembles the actual human behavior~(Fig.~\ref{fig:exp_trajectory}). 

\begin{figure*}
\centering
\includegraphics[width=1.0\linewidth]{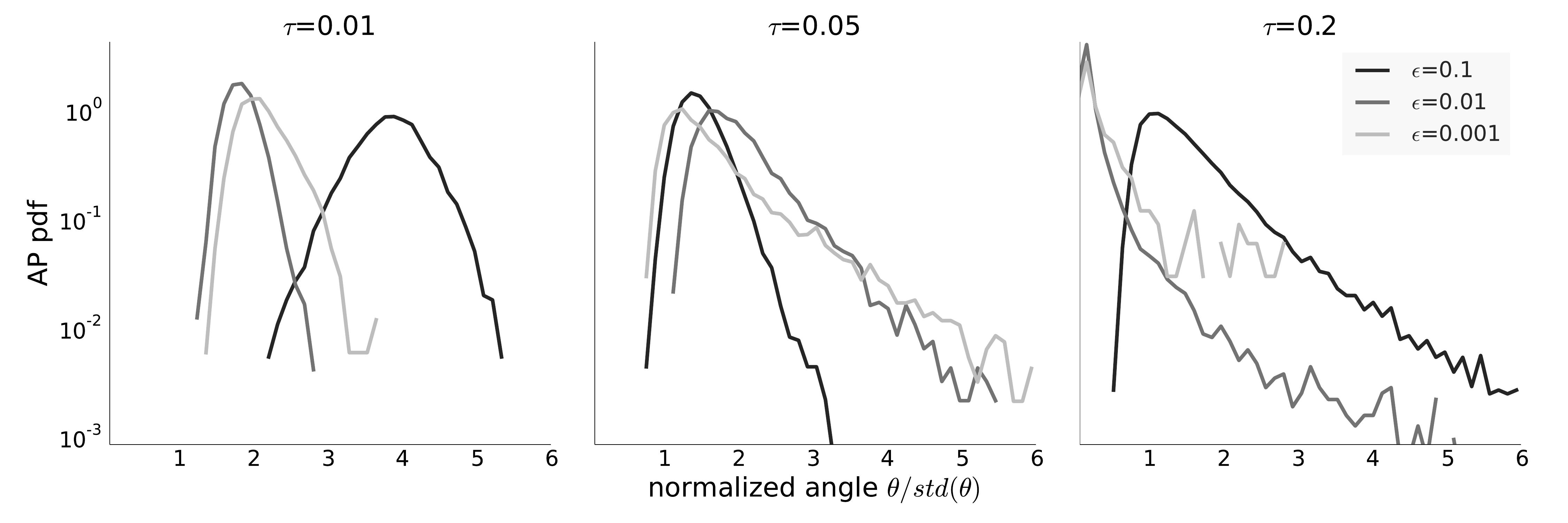}
\caption{
Probability distribution functions (pdf) of absolute values of action points (AP) exhibited by the model for various values of system parameters. The duration of numerical simulation needed to obtain stable distributions is of order $10^6\tau$. Values of $\theta$ are normalized so that the angle values of different scales can be compared.
} 
\label{fig:ap_distr}
\end{figure*}

It should be noted that the amplitude of the stick fluctuations increases with $\tau$ and decreases with $\epsilon$. Indeed, the longer the time $\tau$ of the transient process $\xi:\,0\to1$, the larger the average value reached by the stick angle when the control is finally switched on. On the opposite, increasing the level of noise $\epsilon$ results in higher probability of switching even for small values of $\theta$. These considerations prompt that long $\tau$ together with small $\epsilon$ reflect inability of the controller to switch the control on in response to the steadily increasing $\theta$ (hence large fluctuations in Fig.~\ref{fig:trajectories}, bottom right frame; in the real balancing task the stick would just repeatedly fall in this case).

The distribution of action points (AP; the values of $\theta$ matching the initiation of the transition $\xi:\,0\to1$) highlights the diversity of the control activation dynamics captured by the model (Fig.~\ref{fig:ap_distr}). For short time scales (exemplified by $\tau=0.01$), the log-scale AP distribution has parabolic shape similar to Gaussian, which suggests that in this regime the double-well activation works much like ordinary threshold. When $\tau$ is increased ($\tau=0.05$), small values of $\epsilon$ lead to substantial asymmetry in the AP distribution, which has approximately Laplace-shaped tail part ($\text{pdf} \propto e^{-\theta}$). Finally, if the time scale is further increased ($\tau=0.2$), the exponential shape of the AP distribution tail transforms to the power-law shape with decreasing $\epsilon$.

\begin{figure}
  \centering
  \includegraphics[width=0.8\columnwidth]{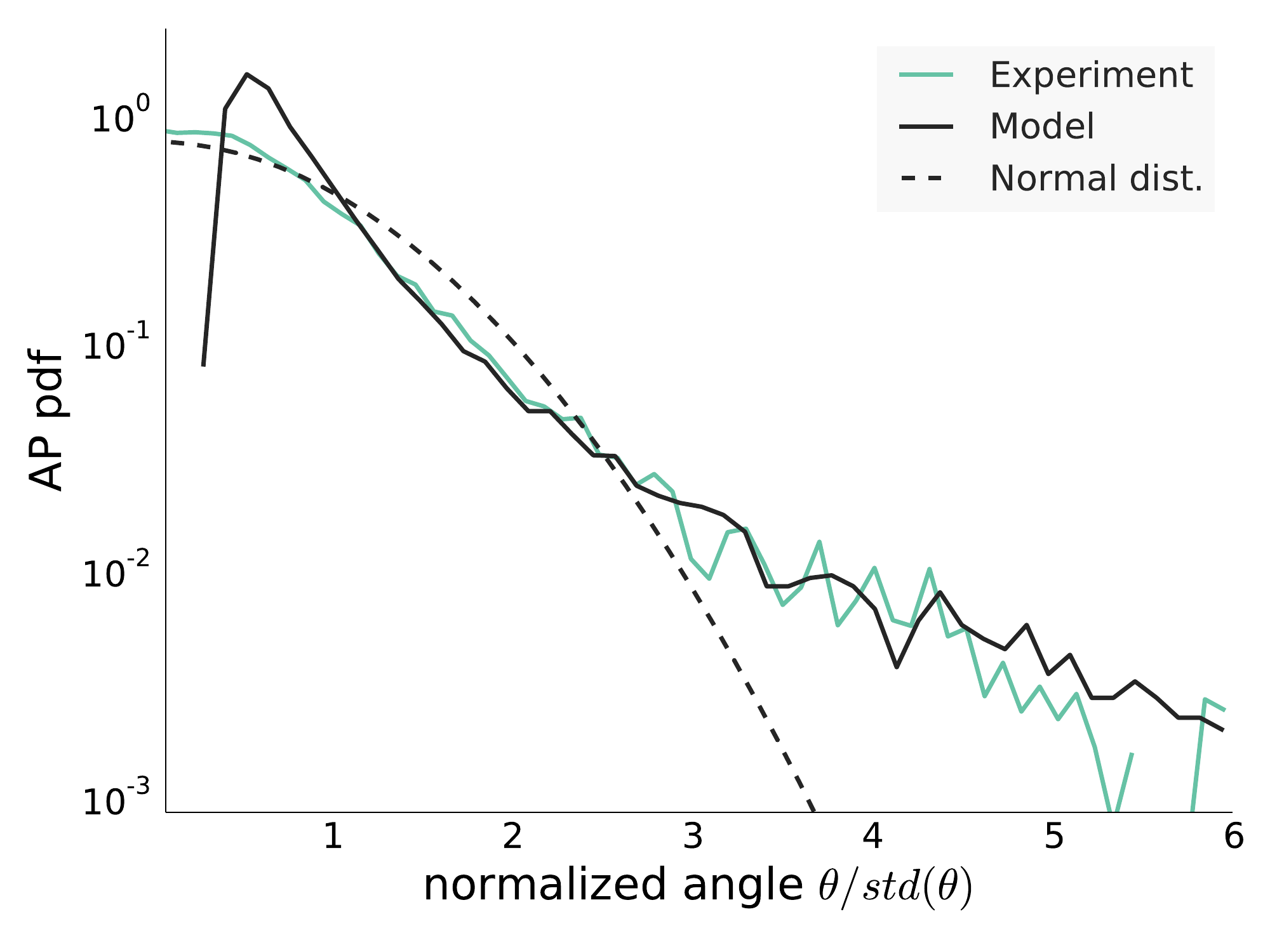}
  \caption{Distributions of action points produced by the model ($\tau = 0.2,\,\epsilon = 0.02$) and human subjects. Experimentally obtained distribution is averaged over ten subjects. Normal distribution truncated at zero is presented for reference.}
  \label{fig:exp_ap_distr}
\end{figure}

Previously obtained experimental data on human overdamped stick balancing revealed distinct Laplace-like distribution of action points~\citep{zgonnikov2014react}. The proposed model captures well the tail part of the experimentally found distribution~(Fig.~\ref{fig:exp_ap_distr}), although for small values of the stick angle the model-generated AP statistics differs from the experimental one. A threshold-based control activation yields the Gaussian distribution of action points (cf. dashed line in Fig.~\ref{fig:exp_ap_distr}), which practically eliminates any probability of control activation at large values of $\theta$. In this sense, noise-driven activation as captured in the double-well model provides a much more plausible explanation of the actual human control strategy.

\section{Discussion}
In controlling unstable systems humans often prefer to switch intermittently between the passive and active behavior instead of controlling the system in a continuous manner. The present paper argues that some intricate properties of intermittent control activation in humans can be explained using the notion of random walk in a double-well potential. We focus on overdamped stick balancing as a representative example of a human control task. In modeling the control activation process, we extend the phase space of the physical stick under human control by an order parameter characterizing the operator's cognitive state. The dynamics of this order parameter is stochastic, and reflects intermittent switching between active and passive behavior of the human operator. The switching process is defined in part by the double-well potential field (changing with the stick angle), and in part by the random force. We describe the latter using multiplicative noise, so that the unambiguous states of the controlled system evoke little noise compared to the uncertain ones. We demonstrate that the model has rich dynamics, and can explain a number of different control activation patterns. By comparing the model to the previously obtained data on human stick balancing, we show that at least one of the modes of the model matches the actually observed human behavior. Based on the presented results, we suggest that employing double-well potential approach to model control activation can aid in understanding complex properties of human intermittent control.

Human brain is a self-organizing system~\citep{haken1996principles,kelso1995dynamic}. Nonlinear interactions of large number of neurons and/or neuronal populations lead to emergence of complex dynamics, which, however, can sometimes be described on the macroscopic level by just a few variables (order parameters). This approach has been widely applied for studying perceptual and cognitive dynamics (see e.g. reviews in~\cite{kelso1995dynamic,rolls2010noisy,wagemans2012century}). At the same time, modern research on human control over unstable systems emphasizes mainly physiological and mechanical aspects of human behavior in diverse balancing tasks, while the underlying cognitive mechanisms are rarely studied. In particular, control activation has been conventionally modeled as a threshold-driven process. However, a question had arisen recently as to whether threshold-based models can fully accommodate complex, often unpredictable dynamics of human-controlled systems~\citep{bottaro2008bounded,cabrera2012stick,asai2013learning}. 

Models implementing noise-driven control activation can potentially serve as richer alternatives to the threshold-based models of intermittent motor control~\citep{zgonnikov2014react}. However, a physically tractable model of noise-driven control activation has been missing up to now, which should necessarily incorporate explicit regulations governing the activation dynamics. The present study develops such a model by adopting the concept of noise-driven switching widely used for describing bistable cognitive phenomena, e.g., as perceptual categorization~\citep{tuller1994nonlinear, vanrooij2002nonrepresentational} and decision making under risk~\citep{vanrooij2013modeling}.

Noise has long been recognized as a key factor in human intermittent control. Particularly, Milton, Cabrera et al.~\cite{cabrera2002onoff,milton2009time,milton2013intermittent} have been developing the idea that state-dependent noise can aid in stabilizing (underdamped) inverted pendulum by forcing the feedback gain to oscillate back and forth across the stability boundary, which turns out to be beneficial for balance control under considerable response delay. Our approach is similar to theirs in that without noise the upright position of the stick is unstable. However, the principal distinction between these approaches is the constructive role of noise, which in our case is to induce switching of the operator's cognitive state between acting and waiting. 

Intrinsic stochasticity of the noise-driven switching allows the model to generate diverse distributions of action points (including Gaussian, Laplacian, and Pareto distributions). This suggests that the double-well model can capture control activation mechanisms of different nature. It remains for future experimental work to investigate whether and under what conditions these different mechanisms are employed by human operators. Moreover, one may speculate that in the single control process human operators may utilize different control activation mechanisms (for instance, for different spatial scales of the controlled system's motion). If this is true, there is need for further modeling efforts. The future models may be based, for example, on the hypothesis that the characteristic time scale of control activation can be state-dependent: when the system approaches the desired position, the time scale may become shorter to enable fine tuning of the system state. This hypothesis may provide an explanation of the found mismatch between the double-well-model- and human-generated action point distributions in the vicinity of the zero angle (Fig.~\ref{fig:exp_ap_distr}).

We wish to underline that the presented model currently utilizes the double-well dynamics to capture not only control activation, but \textit{deactivation} as well. However, the latter process is supposedly determined by the open-loop control execution mechanisms. Given the experimentally observed variability of the corrective trajectories with respect to their end-points (Fig.~\ref{fig:exp_trajectory}), further development of mathematical description of open-loop control is needed to capture the found control deactivation patterns in more detail. This would as well aid in representing the whole spectrum of the possible corrective actions.

We believe that the ideas of the present work may extend beyond the specific field of motor control to a more general class of cognitive processes. Complex, high-dimensional neural systems can exhibit low-dimensional dynamics on the macrolevel~\citep{kelso1995dynamic}. An apt example is the phenomenon of bistable perception, which can be characterized by the double-well dynamics of switching, e.g., between two alternative interpretations of an ambiguous stimulus~\citep{moreno2007noise}. Evidence for double-well stochastic dynamics has also been found in other cognitive processes, for instance, perceptual categorization of speech patterns~\cite{tuller1994nonlinear} and imagined actions~\citep{vanrooij2002nonrepresentational}. Inspired by the results of the present study, we hypothesize that further quantitative investigations may provide grounds for stronger link between such processes and the advanced concepts employed in describing dynamical systems in physics. Particularly, treating the intensity of the neural noise as a state-dependent rather than constant quantity, on the one hand, is physiologically feasible~\citep{lindner2001transmission,silberberg2004dynamics}, and, on the other hand, may potentially enhance the explanatory power of the modern cognitive models of multistable phenomena.

% \bibliographystyle{vancouver}
% \bibliography{library.bib}

\end{document}